\documentclass[journal]{IEEEtran}
\IEEEoverridecommandlockouts
\usepackage{cite}
\usepackage{amsmath,amssymb,amsfonts}
\usepackage{algorithmic}
\usepackage{graphicx}
\usepackage{textcomp}
\usepackage{xcolor}
\usepackage{acronym}
\usepackage{multirow}
\usepackage{array}
\usepackage{adjustbox}
\usepackage{subcaption}
\usepackage{balance}
\usepackage{soul}
\usepackage{url}
\usepackage{comment}

\acrodef{ACF}{AutoCorrelation Function}
\acrodef{ARIB}{Association of Radio Industries and Businesses}
\acrodef{BB-PLC}{Broadband power line communication}
\acrodef{CENELEC}{European Committee for Electrotechnical Standardization}
\acrodef{CEPRI}{China Electric Power Research Institute}
\acrodef{EK}{Evaluation kit}
\acrodef{FCC}{Federal Communications Commission}
\acrodef{KPSS}{Kwiatkowski, Phillips, Schmidt, and Shin}
\acrodef{LV}{Low Voltage}
\acrodef{NB}{Narrowband}
\acrodef{NB-PLC}{Narrowband power line communication}
\acrodef{NFFT}{Nonequispaced Fast Fourier Transform}
\acrodef{NMSE}{Normalized mean squared error}
\acrodef{OFDM}{Orthogonal frequency-division multiplexing}
\acrodef{PLC}{Power line communication}
\acrodef{PRIME}{PoweRline Intelligent Metering Evolution}
\acrodef{PSD}{Power spectral density}
\acrodef{UNB-PLC}{Ultra narrowband power line communication}

\newcolumntype{M}[1]{>{\centering\arraybackslash}m{#1}}

\def\BibTeX{{\rm B\kern-.05em{\sc i\kern-.025em b}\kern-.08em
    T\kern-.1667em\lower.7ex\hbox{E}\kern-.125emX}}
\begin{document}

\title{Long-Term Noise Characterization of \\Narrowband Power Line Communications \\{\Large Measurements, Analysis and Modeling}}

\author{\IEEEauthorblockN{Simone Raponi\IEEEauthorrefmark{2}, Javier {Hernandez Fernandez}\IEEEauthorrefmark{1}\IEEEauthorrefmark{2}, Aymen Omri\IEEEauthorrefmark{1}, and Gabriele Oligeri\IEEEauthorrefmark{2}}\\
\IEEEauthorblockA{\IEEEauthorrefmark{1}Iberdrola Innovation Middle East, Doha, Qatar. \\}
\IEEEauthorblockA{\IEEEauthorrefmark{2}Division of Information and Computing Technology, \\ College of Science and Engineering,\\ Hamad Bin Khalifa University, Qatar Foundation, Doha, Qatar.}}

\maketitle

\thispagestyle{plain}
\pagestyle{plain}

\begin{abstract}

Noise modeling in power line communications has recently drawn the attention of researchers. However, when characterizing the noise process in narrowband communications, previous works have only focused on small-scale phenomena involving fine-grained details. Nevertheless, the communication link's reliability is also affected by long-term noise phenomena that might affect transfer rates at higher layers as well.
This paper addresses the problem of long-term noise characterization for narrowband power line communications and provides a statistical analysis of the long-term trends affecting the noise levels. We present a statistical description of the noise process in the time and frequency domains based on real field measurements in the FCC band ($10$ kHz - $490$ kHz). The collected data comprises more than $1.8$ billion samples taken from three different locations over a time period of approximately 10 days. The noise samples have been statistically analyzed by considering stationarity, autocorrelation, and independence.
Although our results---being unprecedented---are interesting per se, they improve the noise pattern knowledge, thus paving the way for the design and implementation of more robust PLC protocols.
\end{abstract}


\section{Introduction}
\label{sec:introduction}

\ac{PLC} is a technology that exploits the existing power distribution infrastructure to enable communication among devices. Indeed, \ac{PLC} represents one of the enabling technologies that will pave the way to the migration from old power grids (mainly electromechanical infrastructures devoted to the delivery of power from the producer to the consumers) to the modern smart grid paradigm, which involves several new functionalities and services such as smart metering, electricity trading, and transactive energy.

\ac{PLC} features significant advantages with respect to standard communication infrastructures (such as wireless radio, fiber optics, or Ethernet) since it exploits an already existing infrastructure---the one adopted for power delivery---and requires minimal maintenance for communication.

\ac{PLC} is divided into three different categories: \ac{UNB-PLC}, \ac{NB-PLC}, and \ac{BB-PLC}~\cite{StArtPLC,FOR_THE_GRID}. Smart grid, and in particular smart metering, has found in \ac{NB-PLC} a natural partner~\cite{sendin2014strategies,andreadou2016telecommunication,Paper1,Paper2}. \ac{NB-PLC} comprises all the communications that take place in the spectrum between 9 and 500 kHz, with small bandwidth requirements due to the higher noise levels that affect such low frequencies. The noise characterization of \ac{PLC} channels has attracted the attention of many researchers quite recently, and the vast majority of related studies have focused on \emph{small-scale} modeling that involves the analysis of fine-grained phenomena, such as the noise generated by switching home appliances on and off. 

In general, small-scale modeling can be categorized into the following families: (i) colored background noise, (ii) narrow-band noise, (iii) periodic impulsive noise (which can be either synchronous or asynchronous with respect to the main frequency), and finally, (iv) asynchronous impulsive noise~\cite{zimmermann2000analysis}. In general, colored noise and narrow-band noise are considered background noise, while the rest are considered impulsive noise. Impulsive noise is the most important source of error affecting communication links. However, impulsive noise is difficult to predict and it may even be as high as 40 dB above the background noise level~\cite{andreadou2009modeling}. Compared with asynchronous impulsive noise that is dominant in broadband \ac{PLC}, this type of noise contains longer noise bursts that occur periodically with half the AC cycle~\cite{laguna2015use}. 


While small-scale modeling~\cite{degardin2002tce, llano2019access} turns out to be extremely important to guarantee the reliability of communications over \ac{NB-PLC} channels, in this work, we shed light on a completely different aspect, by highlighting the existence of long-term phenomena that can be exploited by higher communication layers to significantly increase the quality of communication. Indeed, driven by the intuition that people's daily activities are characterized by long-term periodic cycles (on the order of hours), we focus our analysis on a larger scale, by considering only the noise power's long-term trends from three different locations.

{\bf Contribution.} This paper introduces the concept of \emph{long-term} noise modeling involving the analysis, characterization, and modeling of the noise process in the \ac{NB-PLC} band over long periods. We collected approximately 10 days of noise power measurements from three different locations in Doha, Qatar, for a total of $> 1.8$ billion samples (publicly available at the following link~\cite{dataset} when the paper will be accepted for publication). We provided a statistical description of different aspects of noise power involving correlation, independence, and stationarity. Finally, we provide some insight into modeling and forecasting noise power over long term periods by exploiting the previous statistical results.

{\bf Paper Organization.} The rest of this paper is organized as follows. Section~\ref{sec:related_work} summarizes important related contributions in the literature. 
Section~\ref{sec:measurement_setup} presents the measurement setup. The different aspects of noise in the frequency domain are detailed in Section~\ref{sec:frequency_domain}. Section~\ref{sec:time_domain} investigates the correlation and the independence of the noise samples. The modeling and forecasting part is presented and detailed in Section~\ref{sec:modeling}. Then, a discussion of the results is presented in Section~\ref{sec:discussion}. Finally, conclusions are drawn in Section~\ref{sec:conclusion}.

\section{Related Work}
\label{sec:related_work}

As mentioned in the introduction, to propose more effective and efficient communication schemes in \ac{NB-PLC} systems there is a need to study the power lines environment, together with the properties and the characteristics of the communication channels. This would allow better exploiting the communication channel, thus increasing the performance of \ac{NB-PLC} systems. Among the distinctive features of power lines, one that is worth mentioning is time-varying non-white noise, which is defined as the sum of noise waveforms produced by appliances that are connected to the power line network~\cite{katayama2006mathematical}. Refer to~\cite{di2011noise} for a brief survey of existing noise models for the in-home \ac{PLC} scenario and to~\cite{shongwey2014impulse} for a brief survey of impulse noise.

A measurement campaign characterizing the non-intentional emissions in the low-voltage section of the electrical grid (within the frequency range of NB-PLC) has been proposed by~\cite{fernandez}. Their analysis highlights tonal emissions by inverters, colored noise from engines and electronic devices, and NB-PLC transmissions replicas. Although the results are related to the same frequency band considered in this paper, the authors did not provide any statistical description of the noise process.

Authors in~\cite{ritzmann} provide a framework for comparing measurement methods in the NB-PLC frequency range by discussing test signals and metrics to be considered.

Another measurement framework for evaluating the emissions of devices in the frequency range 2-150KHz (supraharmonics) is proposed by~\cite{khokhlov}. The authors highlighted how no normative method exists for the measurement of supraharmonics disturbance levels in the grid, and they provide a detailed comparison from existing standards IEC 61000-4-7, IEC 61000-4-30, CISPR 16-1-1 and a critical assessment of their suitability for disturbance measurements in grid applications.

One of the first mathematical models of noise in \ac{NB-PLC} was introduced in~\cite{katayama2006mathematical}. The authors represented noise as a colored cyclostationary Gaussian process whose variance is a periodic time function and whose \ac{PSD} is fitted to the measured noise. The proposed model provides a benchmark for designing and evaluating communication systems under a power line noise environment. 
However, since the model ignores the time-varying spectral behavior of noise, it is inappropriate for \ac{OFDM} systems and only applies to \ac{NB} single-carrier systems.


To overcome this limitation, the authors in~\cite{nassar2012cyclostationary} proposed a cyclostationary noise model for \ac{NB-PLC} that accounts for both the time and the frequency properties of the measured noise. Specifically, the cyclostationarity period of the \ac{NB-PLC} noise is partitioned into several temporal regions, each of them containing the generated noise as a stationary colored Gaussian process. In this way, every region is characterized by a particular \ac{PSD}, that is fitted to the actual noise measurements. Although the proposed model provides a good fitting for the measured \ac{NB-PLC} noise, the authors in~\cite{elgenedy2016cyclostationary} highlighted two severe drawbacks: (i) instead of relying on a mathematical model, the number of stationary temporal regions (together with the regional boundaries) are inferred by visually inspecting the measured noise spectrogram; and (ii) the noise process within each region is carried out independently of the other regions, thus ignoring any possible cross-correlation between the different noise processes across the regions. To address these drawbacks, in the same study, the authors synthesized the \ac{NB-PLC} noise samples by relying on a frequency-shift filter, which was designed to shape an input white noise spectrum extracted from the experimental noise measurements. Even by keeping the same computational complexity of~\cite{nassar2012cyclostationary}, the model proposed in~\cite{elgenedy2016cyclostationary} exhibited a performance gain in terms of the \ac{NMSE} in the cyclic autocorrelation.

In~\cite{andreadou2009modeling}, the authors proposed a simulated noise model that describes the noise affecting a communication system using the \ac{OFDM} modulation of a power line channel. The proposed model, which is easily implementable in computer simulations, takes account of the component parameters' statistical properties and depicts the features of both the background and the impulsive noise.

A statistical model for the asynchronous impulsive noise in \ac{PLC} networks was put forward in~\cite{nassar2011statistical}. The authors first exploited the physical properties of the \ac{PLC} network to derive a statistical-physical model of the instantaneous statistics of asynchronous noise, and then validated the distribution by relying on both simulated and measured \ac{PLC} noise data. Impulsive noise in (indoor) \ac{NB-PLC} was also studied in~\cite{gassara2015novel}, in which the authors proposed a stochastic model to provide an accurate characterization. They first extracted both the pulses and their characteristic parameters, composing four empirical classes.
Then, they determined which distributions most effectively fit the data. According to the study: the pulses pseudo-frequency deduced from the measurements followed a log-normal distributions at three different intervals; the peak amplitude was well fitted by Beta distributions, the duration by Weibull distributions, while the interarrival time was lognormally distributed. Furthermore, the normalized damping factor of the sinusoids follows Gamma distribution, and the phase of the pulses appeared to be normally distributed by absolute value.

The use of the $\alpha$-model as an alternative to the Gaussian model to capture \ac{PLC} background noise was proposed by~\cite{laguna2015use}. The authors provided evidence that the marginal distribution of the noise found in \ac{PLC} systems exhibits particular statistical properties that can be well modeled by relying on an $\alpha$-stable distribution. After presenting a practical application of the model to synthesize noise in the power line, the authors evaluated the performance of an \ac{OFDM} communication system under both Gaussian and  $\alpha$-stable background noise. Among their conclusions, they pointed out that, when highly impulsive background traffic is (mistakenly) assumed as Gaussian in evaluation models, there is a risk of largely overestimating the performance of the system. It is worth mentioning that the same authors had already used the $\alpha$-model to describe both the impulsive and the background noise for \ac{PLC} systems in industrial environments~\cite{laguna2014experimental}.

In addition to the Gaussian model and the $\alpha$-model, many other distributions have been introduced to fit the noise data in \ac{PLC}. Among them, Middleton's class A distribution~\cite{umehara2004turbo}, the Nagakami-m distribution~\cite{meng2005modeling}, and the Rayleigh distribution~\cite{chan1989amplitude} are worth mentioning.

In~\cite{cortes2016suitability}, the authors focused on Middleton's class A distribution and analyzed its suitability to model noise amplitude in NB-PLC. The analysis was carried out by relying on 311 noise registers measured in the CENELEC-A band. According to the study, Middleton's class A distribution exhibits a quite limited capability to represent the measured noise since only a small percentage of the registers (i.e., 14.47\%) presents amplitudes drawn from the distribution.

\section{Measurement Setup} 
\label{sec:measurement_setup}

Our noise field measurements were performed in the \ac{NB-PLC} \ac{FCC} band ($10$ kHz - $490$ kHz) and were segmented into eight sub-bands, following the \ac{PRIME} standard version 1.4~\cite{PRIME14}. The eight channels were within the frequency band from $42$ kHz to $471$ kHz, where the bandwidth of each channel and the sub-carrier mapping were presented according to \ac{PRIME} protocol~\cite{PRIME14}. 

An evaluation kit for the PL360 modem from Microchip~\cite{PL360HC,PL360UG} was used to perform the long term \ac{NB-PLC} noise field measurements. The PL360 is a multi-protocol \ac{PLC} modem designed with a flexible architecture that allows the implementation of standard and customized \ac{PLC} solutions compliant with CENELEC EN50065, FCC, and ARIB regulations~\cite{PL360HC}. 


The PL360 modem features a sampling frequency of $1$ MHz with $NFFT=2048$ samples, which results in a frequency resolution, or sub-carrier spacing, $\Delta f = 488.28$ Hz as also defined by PRIME technology. Therefore, every $1$ s (noise data sampling period), PL360 modem logs 2048 samples with a signal sampling frequency of $1$ MHz (one sample every $1$ us) and a resolution of 16 bits. It is worth noting that the PRIME channels' maximum frequency is still under the Nyquist–Shannon threshold, i.e., 471 kHz $<$ 1 MHz / 2.  Finally, PRIME defines $8$ channels, with 97 sub-carriers with frequency spacing $\Delta f = 488.28$ Hz, thus the modem logs the noise levels for the different sub-carriers within each channel by using the corresponding indexes from the noise vector of 2048 samples. We highlight that, in this study, we do not consider the 15 subcarriers $\times$ 7 guard intervals between adjacent channels.

\begin{figure}[t]
    \centering
    \includegraphics[width=0.5\columnwidth]{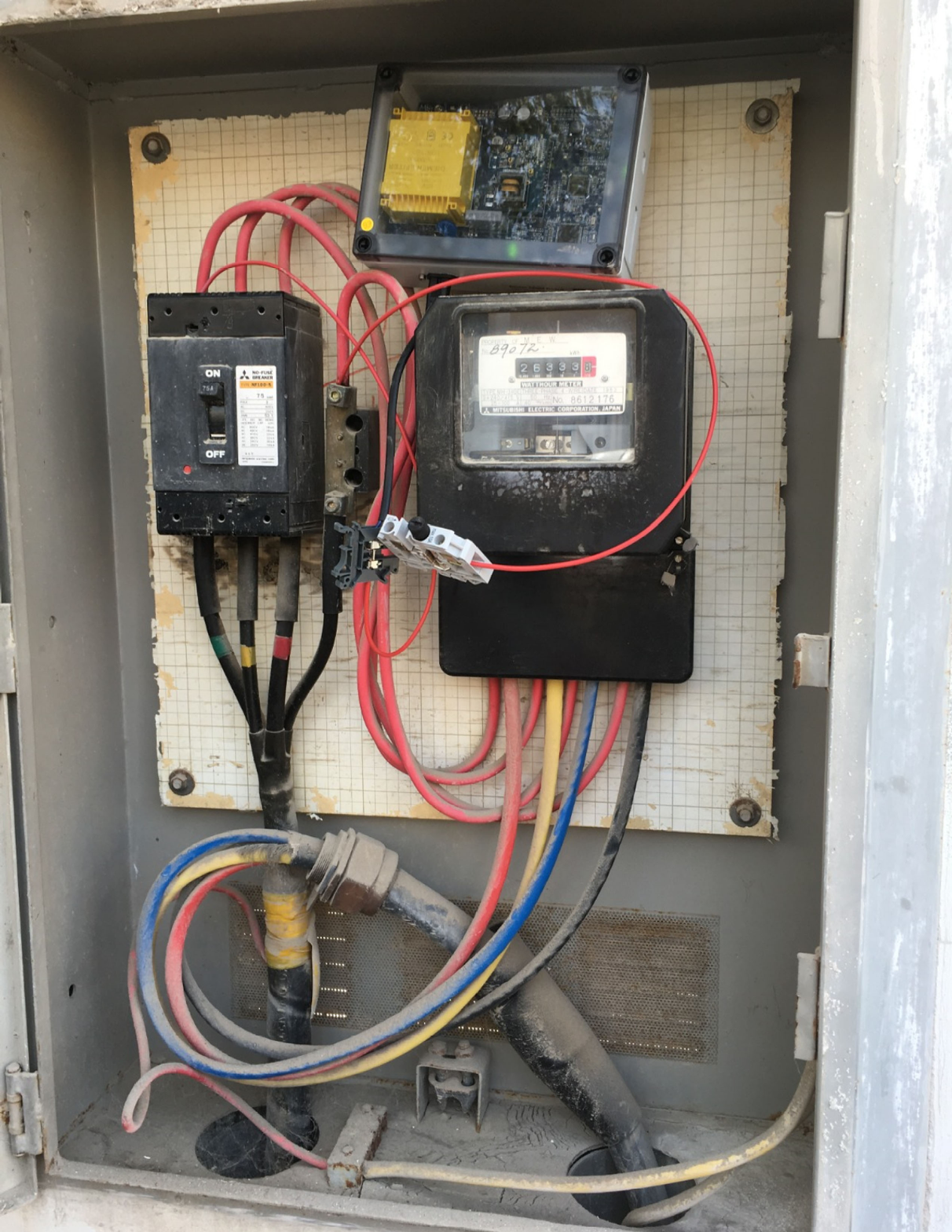}
    \caption{\ac{PLC} modem coupled to a \ac{LV} line at the customers meter distribution box.}
    \label{fig:microchip_evaluation_kit}
\end{figure}

{\bf Locations.} The measuring devices were directly connected to the \ac{LV} lines as close as possible to the electrical meter. Fig.~\ref{fig:microchip_evaluation_kit} shows a \ac{PLC} modem coupled to a \ac{LV} three-phase power meter for gathering data. Nevertheless, we observe that our measurements were limited to a single phase. A laptop, connected via USB to the measuring device, was used to store the data. The measurements were acquired from three customers' premises, i.e., Locations 1, 2, and 3 as per Table~\ref{table:locations}, which vary in type (i.e., apartment or villa), size (from 100 to 260 $m^2$), and covered Qatar's three most common areas: semi-rural, high density urban, and low density urban.

{\bf LV distribution grid.} Qatar's distribution network has a rated voltage of 240/415V, neutral solidly earthed, and nominal mains frequency of 50 Hz. LV lines run underground with an average length of 350-400 meters. The average number of customers per transformer station spans between 50 to 75 in high-density areas while from 10 to 15 in low density areas~\cite{kah2018}.

\begin{table}[]
\caption{Description of the Locations involved in our measurement campaign}
\centering
\begin{adjustbox}{width=\columnwidth}
\begin{tabular}{l|cccc|}
 & \textbf{Type} & \textbf{Area ($m^2$)} & \textbf{N. of people} & \textbf{Environment} \\ \hline
L1 & Apartment & 100 & 4 & Urban (High density) \\
L2 & Apartment & 260 & 3 & Urban (Low density) \\
L3 & Villa & 250 & 1 & Semi-rural \\ \hline
\end{tabular}
\end{adjustbox}
\label{table:locations}
\end{table}

{\bf Sampling period.} Our measurement campaign lasted for 10 days at different locations. The measurements were recorded in dB$\mu$V unit, with a sampling period of $1$ second during 10 days. Moreover, every $1$ second, the PL360 kit collects 2048 noise samples (in the frequency range between $42$ kHz and $471$ kHz) at the different sub-carriers with a sampling time of $1$ $\mu$s, corresponding to a sampling frequency of $1$ MHz.  We stress that our analysis aimed at highlighting the long-term phenomena of the noise process, i.e., involving minutes or even hours. Contrary to other contributions, we are not aiming to model small scale effects lasting for $\mu s$, such as impulsive noise, but we are interested in investigating how the day-to-day human behavior and household appliances affect the noise process on a long-term time frame.

 \begin{figure}[t]
    \centering
    \includegraphics[width=0.80\columnwidth]{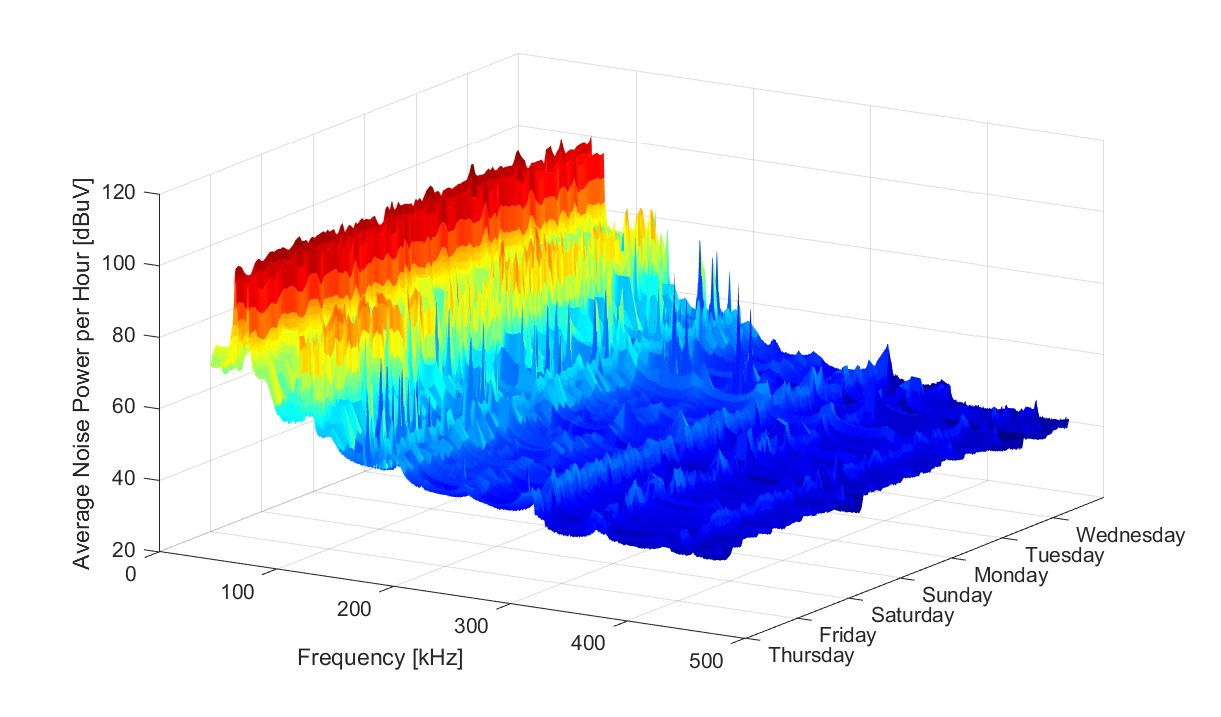}
    \caption{NB-\ac{PLC} noise power variation during one week at Location 2.}
    \label{fig:NoiseVar3D}
\end{figure}

\begin{figure}[h]
    \includegraphics[width=0.8\columnwidth]{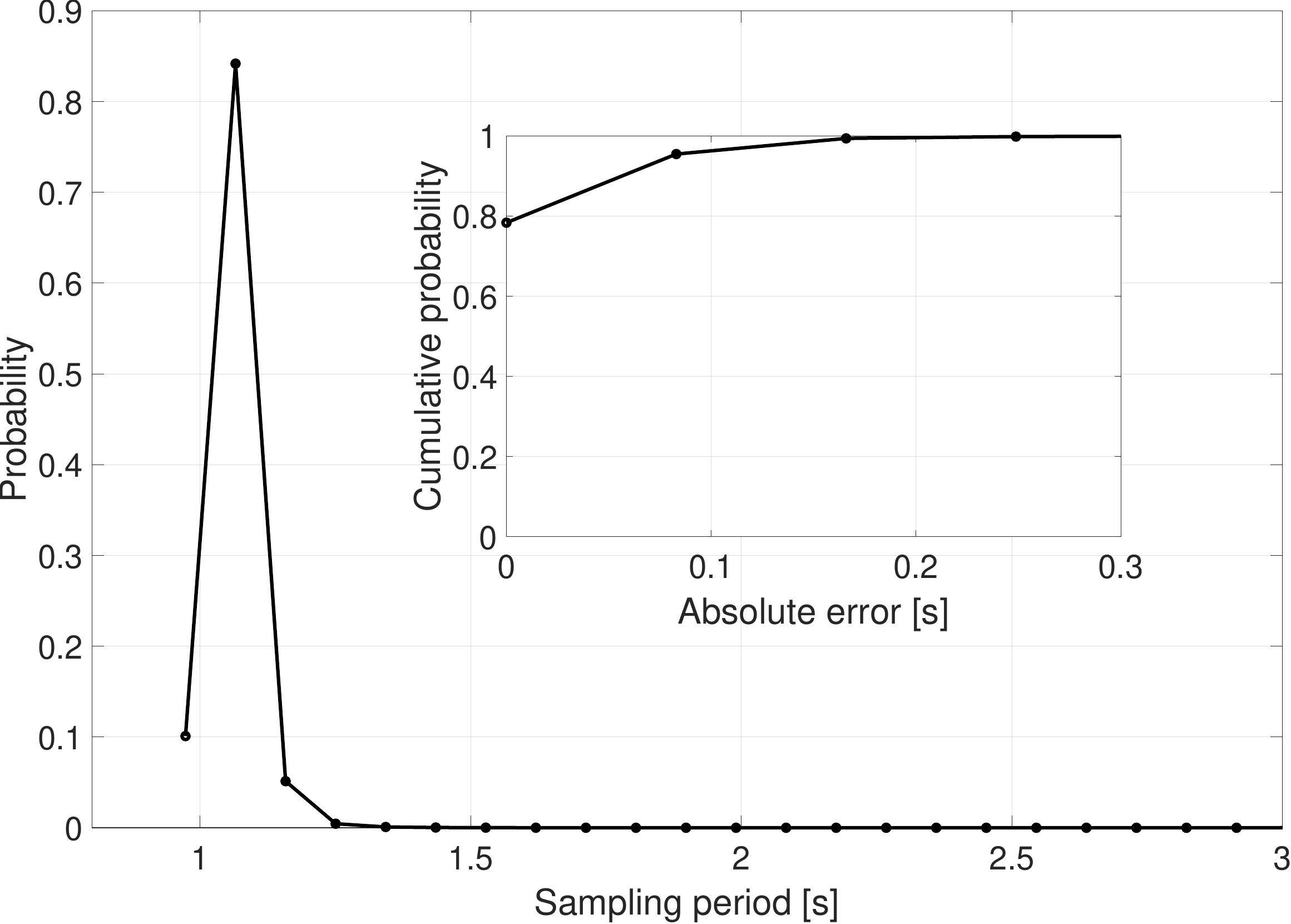}
    \centering
    \caption{Probability distribution function associated with the actual sampling period and cumulative probability associated with the absolute error. The sampling procedure is affected by only minor delays, i.e., less than 65ms respect to the requested sampling of 1 sample per second.}
    \label{fig:sampling_analysis}
\end{figure}

We performed a preliminary analysis of the collected noise values that highlighted the quality of the sampling process. Figure~\ref{fig:sampling_analysis} shows the probability distribution function associated with the time-gap between two consecutive samples. The analysis was performed considering all the noise values from all three locations for a total of 1,826,665,198 samples. We observed a sampling period of 1s recorded for 1.0655, which introduced an error (delay) of approximately 65ms. We observed other spurious sampling periods that were statistically insignificant as proved by the cumulative distribution function in the inset figure of Fig.~\ref{fig:sampling_analysis}, i.e., 95\% of the sampling periods had an absolute error of less than 0.08 s.

\section{Frequency Domain}
\label{sec:frequency_domain}

In this section, we consider different aspects of the noise in the frequency domain.

\begin{figure*}[t]
    \minipage{0.32\textwidth}
        \includegraphics[width=\linewidth]{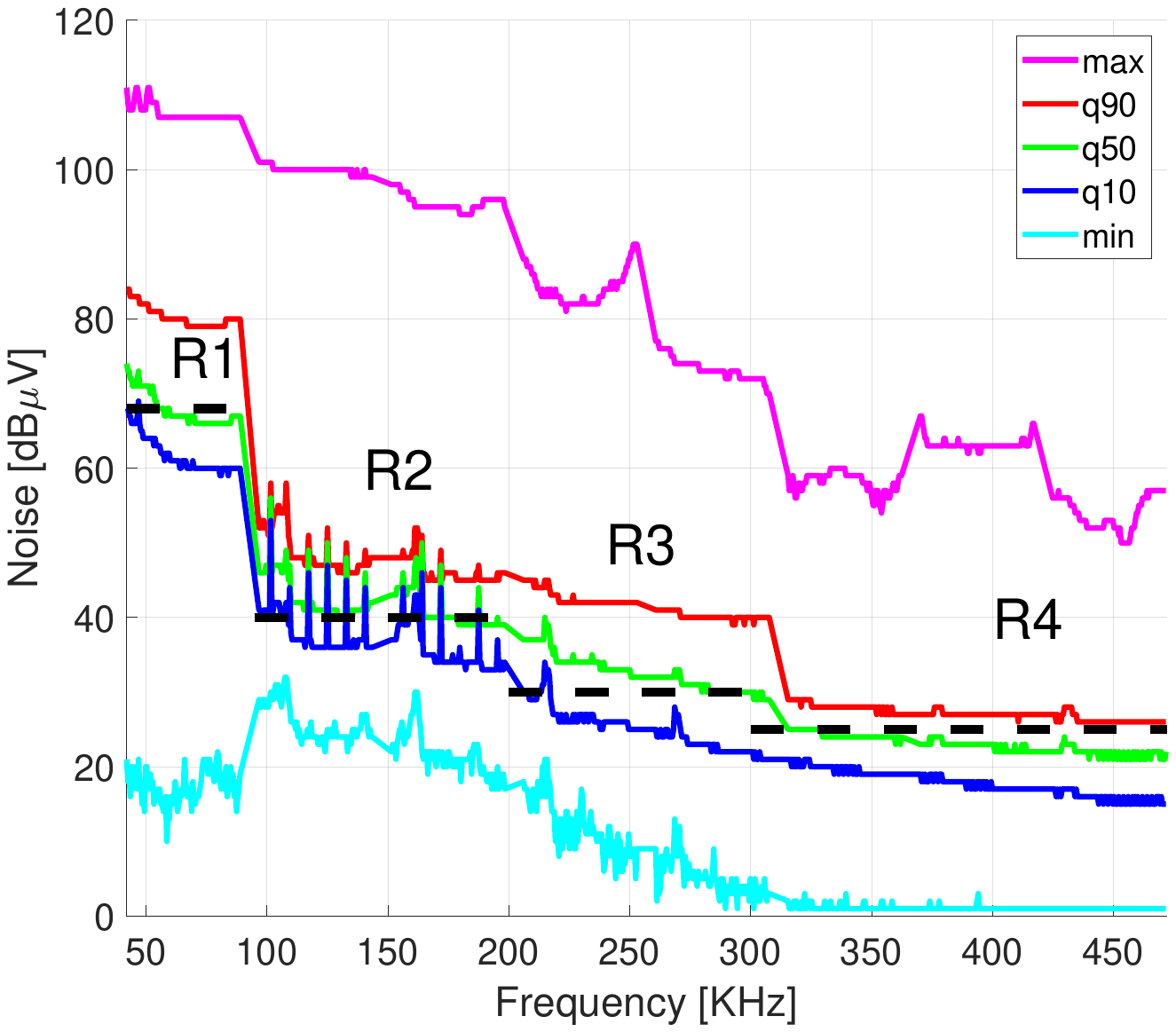}
        \subcaption{Location 1}\label{fig:spectrum_aymen}
    \endminipage\hfill
    \minipage{0.32\textwidth}
        \includegraphics[width=\linewidth]{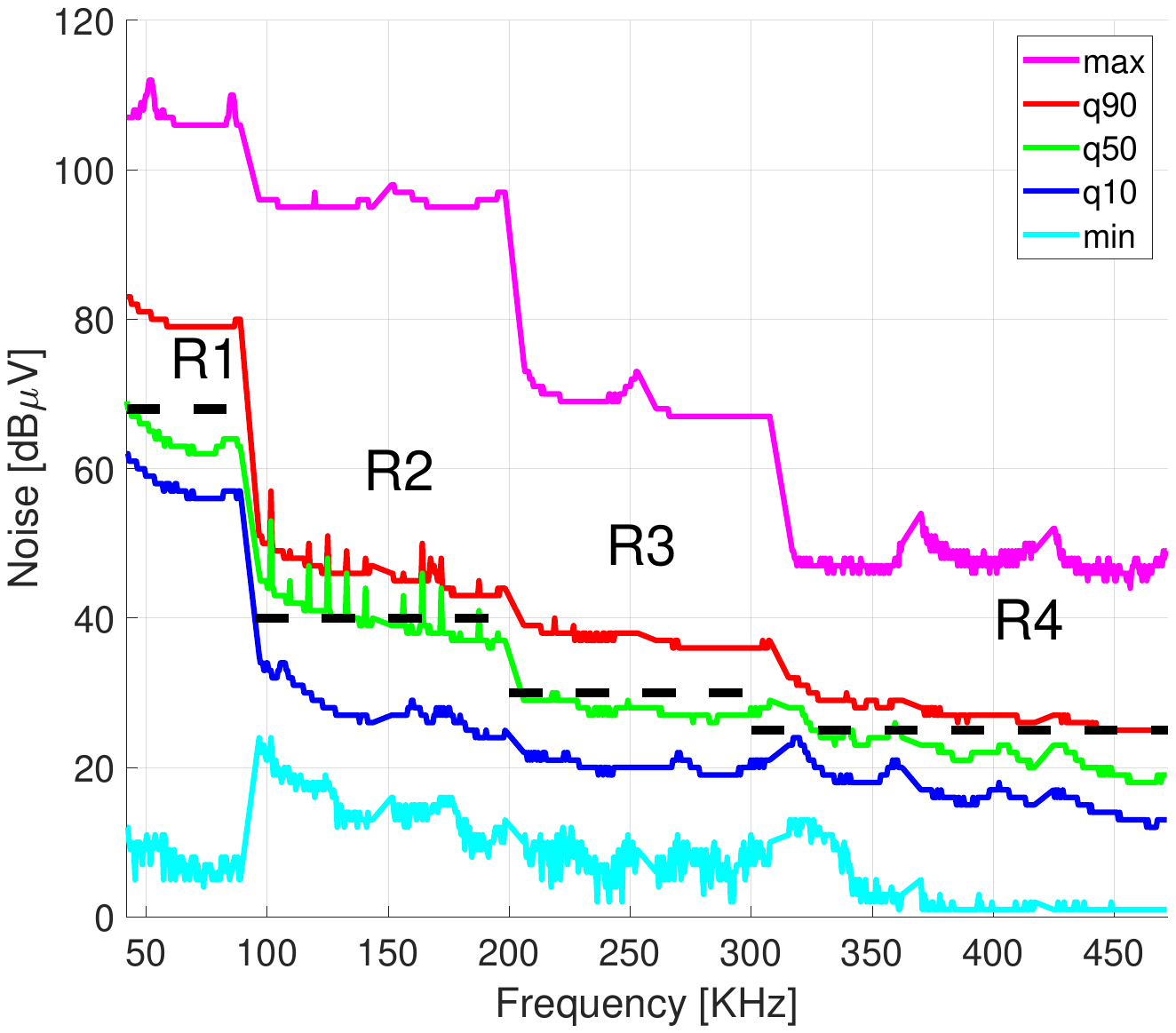}
        \subcaption{Location 2}\label{fig:spectrum_javier}
    \endminipage\hfill
    \minipage{0.32\textwidth}%
        \includegraphics[width=\linewidth]{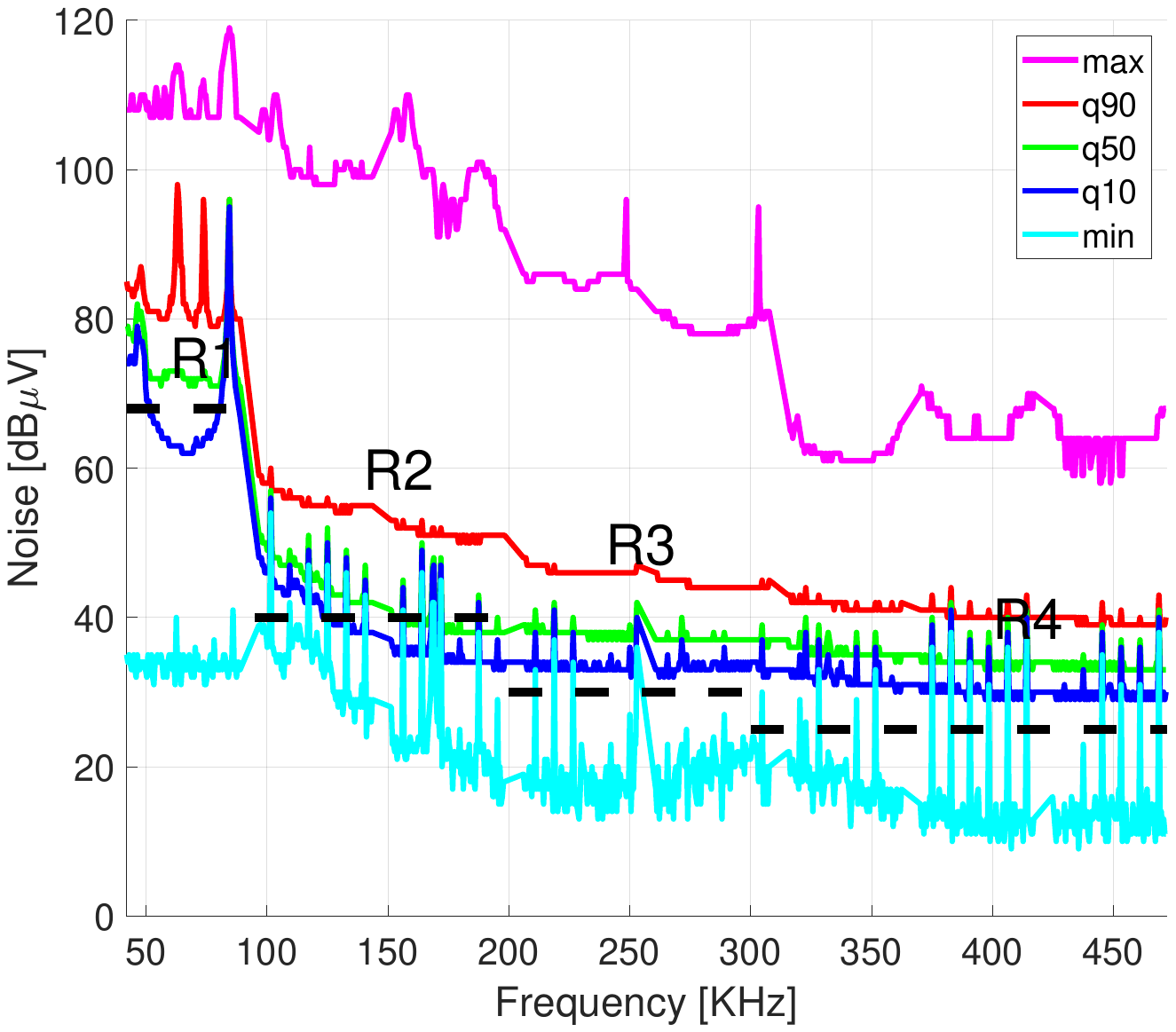}
        \subcaption{Location 3}\label{fig:spectrum_gabriele}
    \endminipage
    \centering
    \caption{Noise levels ($dB\mu V$) as a function of the frequency (kHz). We considered three different locations (1, 2, and 3) and more than one week of noise samples taken at 1 sample per second (1Hz). We computed minimum (cyan), maximum (magenta), and quantile 10 (blue), 50 (green), and 90 (red). The dashed black line highlights the four regions (R1, R2, R3, and R4) with similar noise characteristics.}
    \label{fig:spectrum}
\end{figure*}

{\bf Spectrum-wise analysis.} Our preliminary analysis consisted of estimating the noise levels ($dB\mu V$) as a function of the frequency spanning between 41,992 kHz and 471,679 kHz. For each of the 776 frequencies (97 subcarriers $\times$ 8 channels), we considered a time series of 639K measurements of the noise (more than one week), and we computed the minimum, quantiles 10 (q10), 50 (q50), 90 (q90), and finally, the maximum. Figure~\ref{fig:spectrum} shows the statistics mentioned above as a function of the considered frequencies. First, we observe that lower levels of noise characterize higher frequencies, i.e., there is a clear trend---involving all the statistics---according to which the level of noise decreases when the frequency increases. We identify the following regions:

 \begin{itemize}
    \item {\bf Region 1 (R1)} [41.9 kHz - 95 kHz]. In this region, the noise is almost independent of the frequency and characterized by a median value equal to 68 $dB\mu V$. The difference between quantile 90 and quantile 10 ($q90 - q10$) is approximately 20 $dB\mu V$  across the region, while the overall variance (max-min) sums to $90 dB\mu V$. This means that the 80\% of the samples are in the tight range of $\approx$ 20 $dB\mu V$, while there are a few (less than 20\%) samples characterized by high variance, i.e., $\approx \pm$30 $dB\mu V$.
    \item {\bf Region 2 (R2)} [95 kHz - 200 kHz]. As in the previous case, the noise is (almost) independent of the frequency and is characterized by a median value of approximately $40 dB\mu V$, while the q90-q10 variance is smaller than $20 dB\mu V$.
    \item {\bf Region 3 (R3)} [200 kHz - 300 kHz]. A lower level of noise characterizes this region with respect to R1 and R2, i.e., approximately 30 $dB\mu V$ considering the median value (solid green line). However, we observe that Location 3 is affected by higher noise values, i.e., approximately $39 dB\mu V$ (median value). The overall noise variance (q90 - q10) is quite small, i.e., 15 $dB\mu V$, while the min-max variance is significantly smaller than in previous cases, being at approximately 60 $dB\mu V$, which is mainly due to an increased minimum level of noise.
    \item {\bf Region 4 (R4)} [300 kHz - 471.68 kHz]. This region is characterized by  flat noise with a median value of approximately 23 $dB\mu V$. The noise variance is about 10 $dB\mu V$, while the min-max distance is the same as the one in Region 2 ($\approx 65 dB\mu V$). Finally, we observe a few anomalies in the maximum levels of noise with a period of 55 kHz, with spikes of 22 $dB\mu V$ of amplitude.
\end{itemize}

\begin{figure}[h]
    \includegraphics[width=0.8\columnwidth]{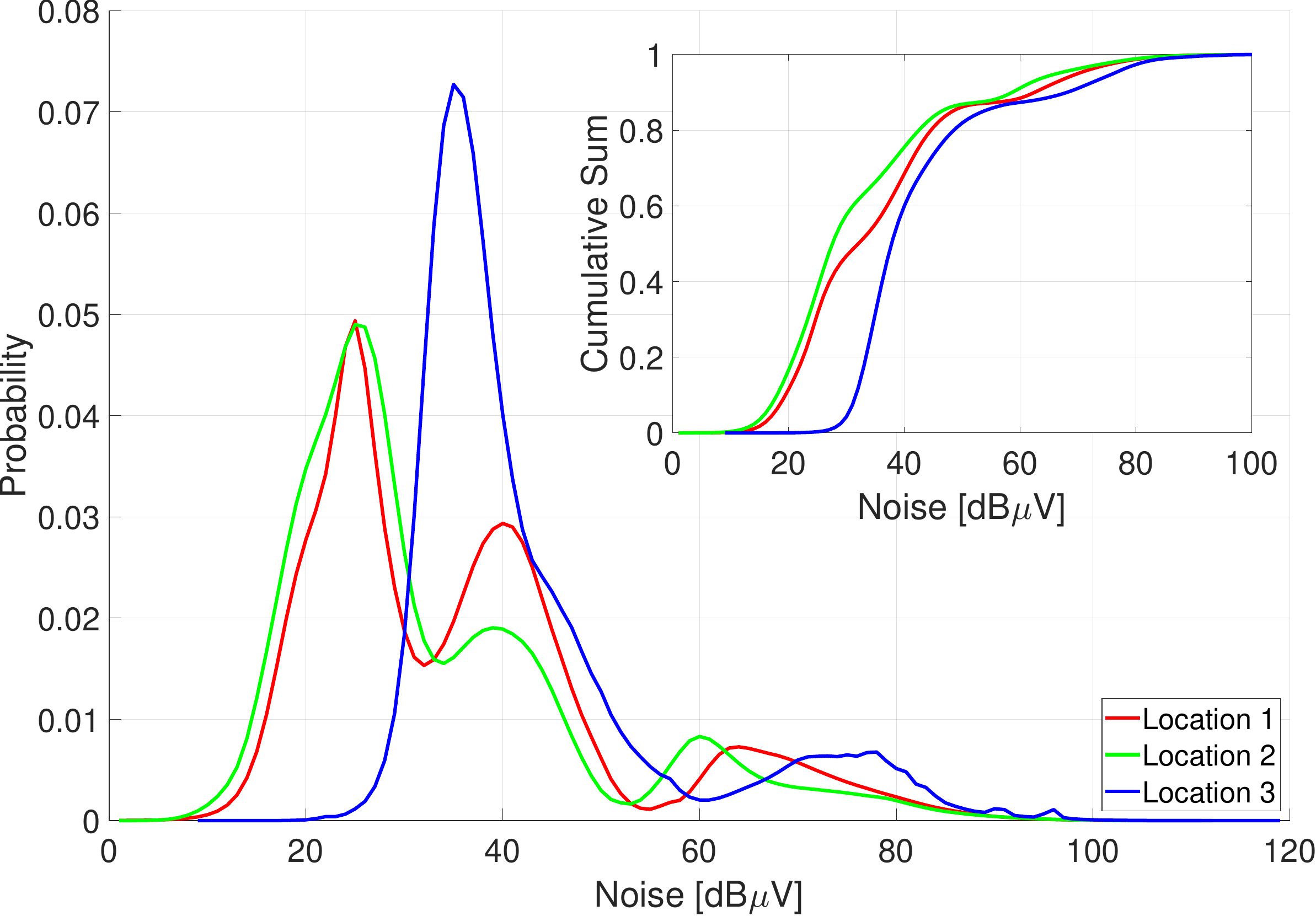}
    \centering
    \caption{Probability distribution function associated with all the noise sample values. The inset figure refers to the cumulative distribution function associated with the noise values.}
    \label{fig:all_noise_distribution}
\end{figure}

Figure~\ref{fig:all_noise_distribution} shows the probability distribution function associated with all the noise values, i.e., considering all samples for all frequencies. The three peaks at 25, 40, and 65 $dB\mu V$ match the average values of the previously introduced regions. Finally, we observe that 80\% of the noise values (q90 - q10) are in the range between 20 and 64 $dB\mu V$ (inset figure of Fig.~\ref{fig:all_noise_distribution}).

\section{Correlation and Independence of the Noise Samples}
\label{sec:time_domain}

This section provides an in-depth analysis of the stationarity and independence of the collected noise samples by taking different tests and statistics into account.

\begin{figure}[h]
    \includegraphics[width=0.80\columnwidth]{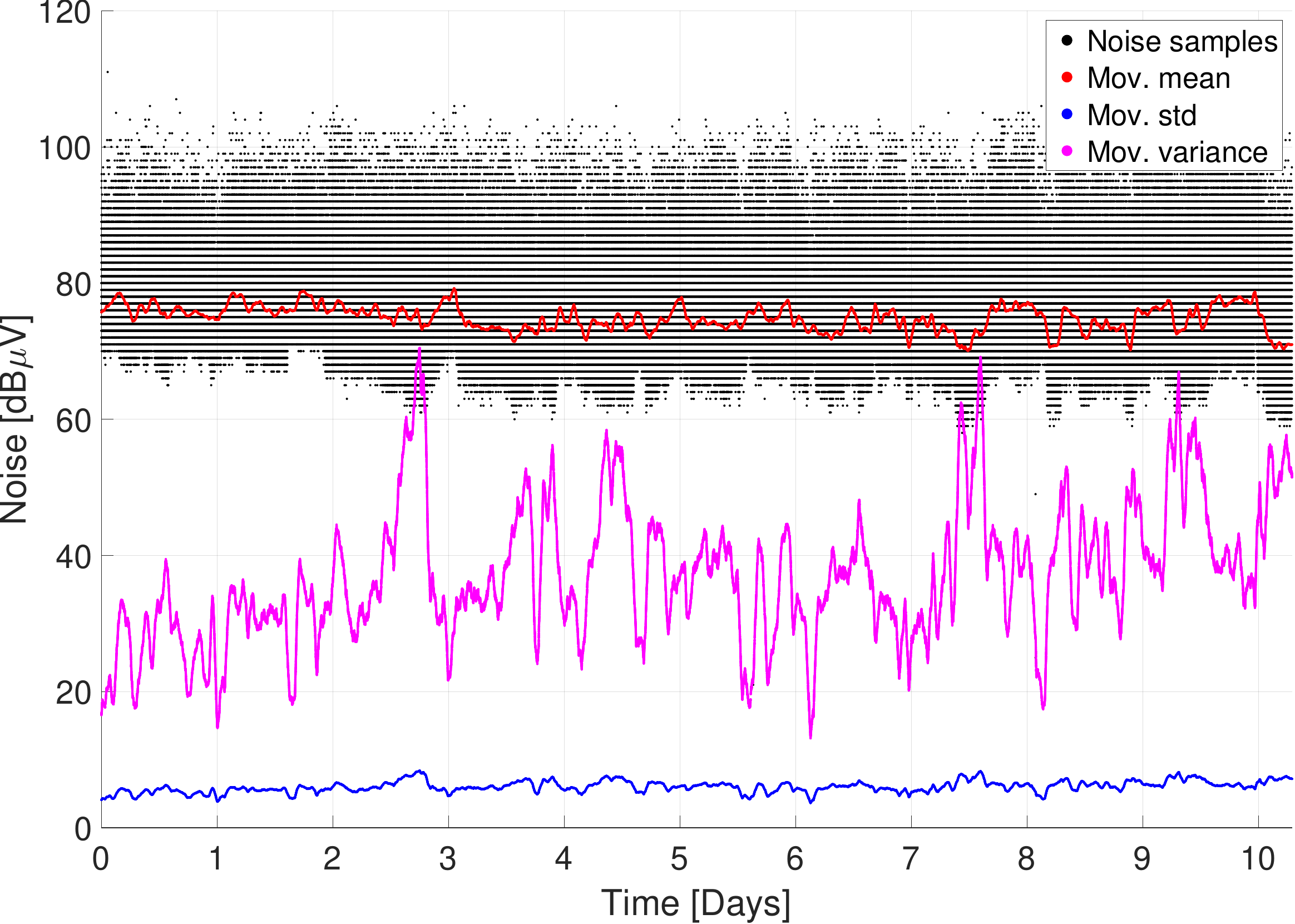}
    \centering
    \caption{First/Second-order statistics (moving mean and standard deviation with window size of 1 hour) computed over the noise values of frequency 1.}
    \label{fig:noise_time}
\end{figure}

{\bf Stationarity.} We start our analysis by considering first-order statistics. Figure~\ref{fig:noise_time} shows a typical sequence (Frequency 1) of noise values (black dots) of approximately 10 days. We considered the mean (red dots), the standard deviation (blue dots), and the variance (magenta dots) over a sliding window of 3600 samples (1 h). In particular, Fig.~\ref{fig:noise_time} refers to all the samples collected from Location 1 - Frequency 1. Due to space constraints, we do not report the same analysis for all the locations and frequencies. Nevertheless, we carefully went through all the data samples and we confirm that all of them are characterized by similar trends.

The objective of our analysis was to evaluate the stationarity of the random process associated with the noise. We started our analysis by considering first and second-order statistics, i.e., the mean and variance from Fig.~\ref{fig:noise_time}, which were characterized by minute fluctuations. Conversely, the raw values of the noise were affected by large fluctuations between 13 and 70 $dB\mu V$, as depicted by the moving variance in Fig.~\ref{fig:noise_time}. Given the considerations mentioned above, our intuition was that the noise values represented a stationary sequence. 
We used the \ac{KPSS} test for a unit root in the time series~\cite{kpsstest}. Our analysis considers one subcarrier from each channel and adjacent chunks of different sizes ranging between 30 and 600 samples (0.5 min and 10 min, respectively). For each chunk, we ran the \ac{KPSS} test, which gave 0 to indicate we should reject the stationary in favor of the unit root alternative, and 1 when the test failed to reject the stationarity. We evaluated the fraction of stationary chunks for each chunk size by summing the chunks with a test output of 1 with respect to the total number of chunks. 

\begin{figure*}[h]
    \minipage{0.33\textwidth}
        \includegraphics[width=\linewidth]{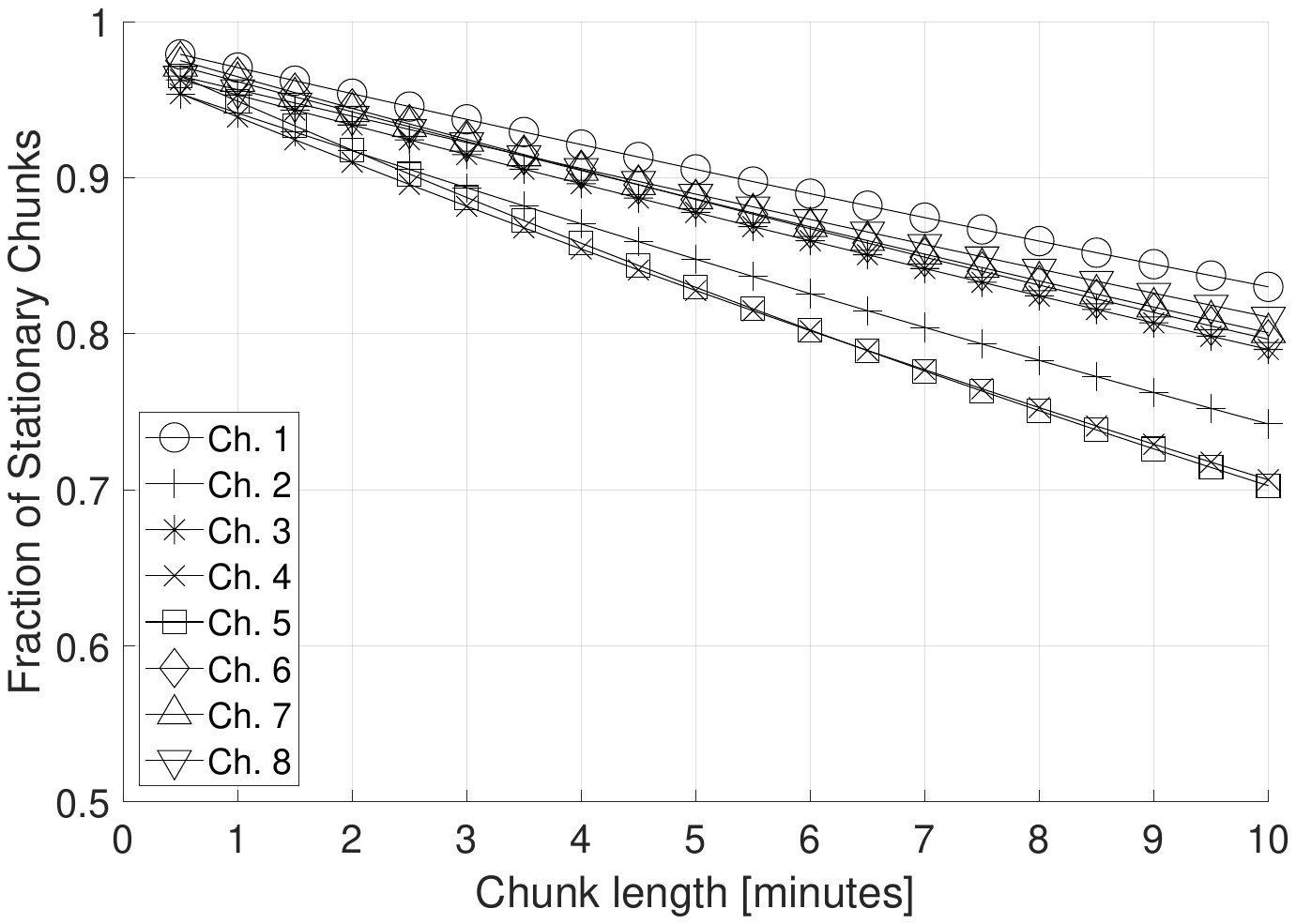}
        \subcaption{Location 1}\label{fig:stationarity_aymen}
    \endminipage\hfill
    \minipage{0.33\textwidth}
        \includegraphics[width=\linewidth]{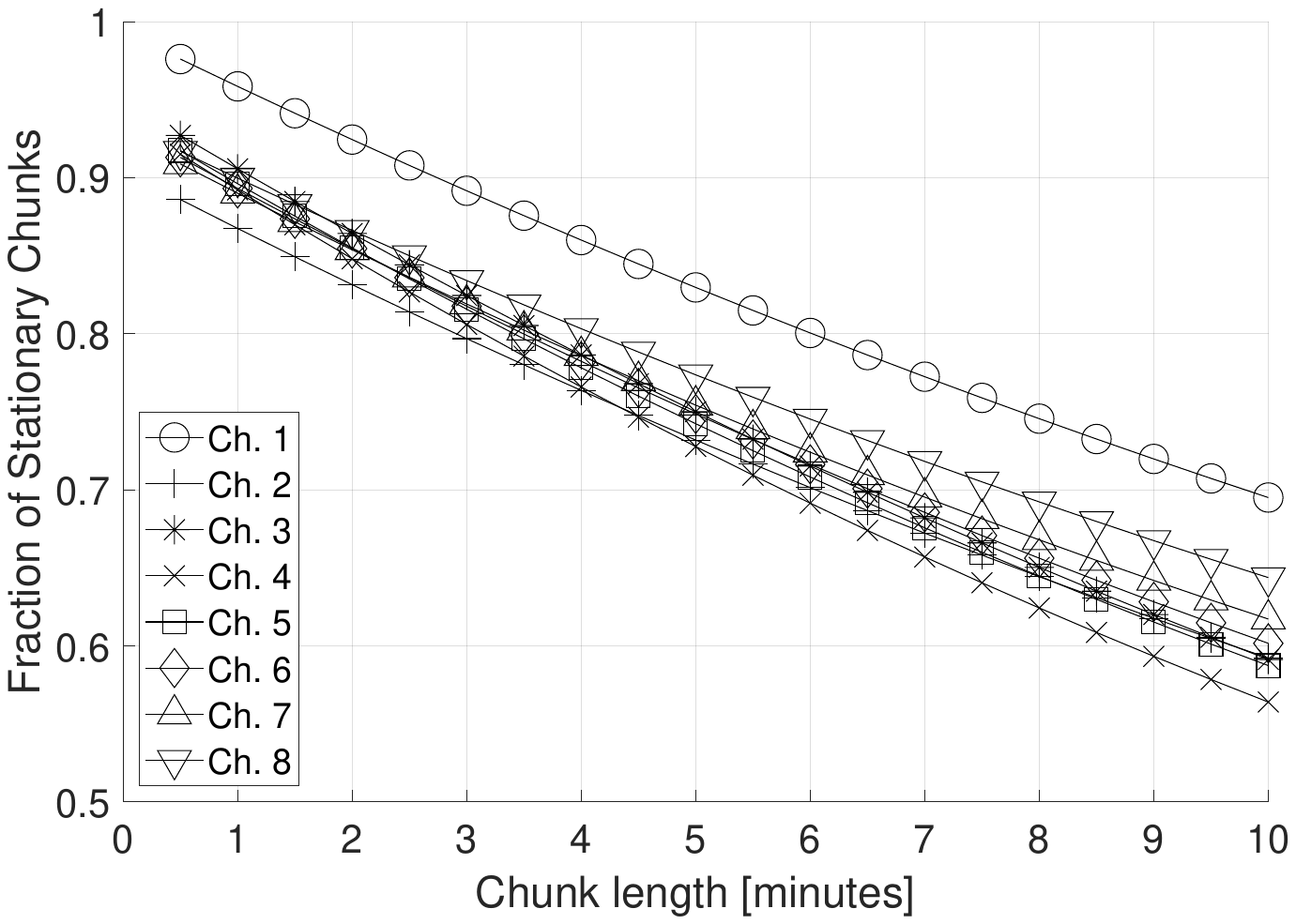}
        \subcaption{Location 2}\label{fig:stationarity_javier}
    \endminipage\hfill
    \minipage{0.33\textwidth}%
        \includegraphics[width=\linewidth]{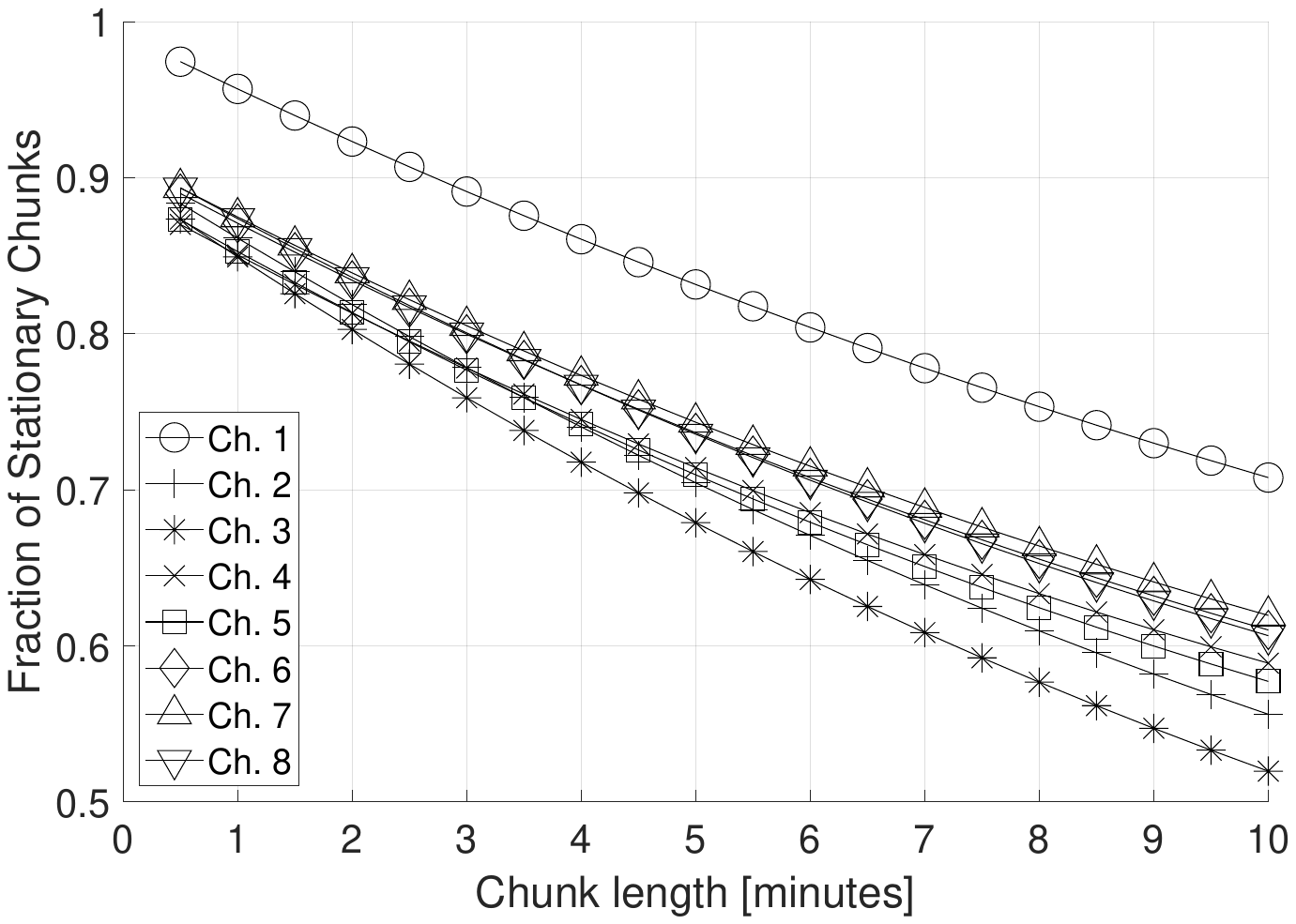}
        \subcaption{Location 3}\label{fig:stationarity_gabriele}
    \endminipage
    \centering
    \caption{Fraction of stationary chunks as a function of chunk length and the three considered locations.}
    \label{fig:stationarity}
\end{figure*}

Figure~\ref{fig:stationarity} shows the results of our stationarity analysis considering the three locations. The fraction of chunks that are positive for the stationarity test decreases when the chunk's length becomes larger, i.e., larger chunks are more likely to contain non-stationary sub-sequences. The two chunk extremes, i.e., 0.5 and 10 min, have similar values at locations 2 and 3, while location 1 is characterized by higher stationarity levels for larger chunks. It is worth noting that the noise process presents a high level of stationarity for long periods of time: chunks shorter than 2 min (120 samples) in Location 1 are stationary in the 90\% of the cases while, to have similar values in Location 2 and 3, chunks length should be reduced to approximately 30 seconds (30 samples). This result is quite significant because it implies that the noise process is stationary for long periods of time, i.e., between 30 s and 2 min depending on the considered location.

Finally, we performed a similar analysis with the augmented Dickey-Fuller test and the Phillips-Perron test for one unit root, and we obtained similar results.

\begin{figure*}[h]
    \minipage{0.33\textwidth}
        \includegraphics[width=\linewidth]{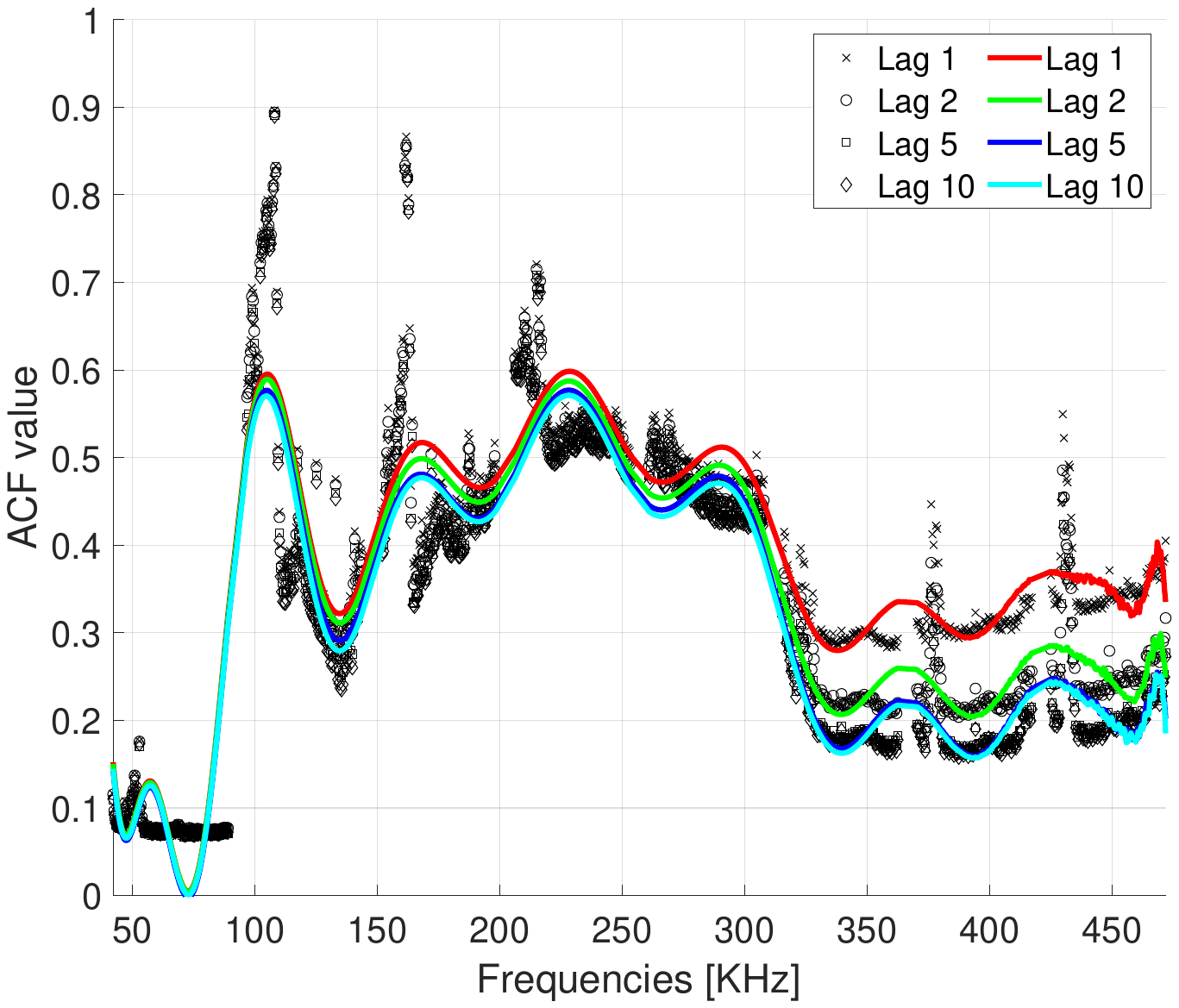}
        \subcaption{Location 1}\label{fig:lags_aymen}
    \endminipage\hfill
    \minipage{0.33\textwidth}
        \includegraphics[width=\linewidth]{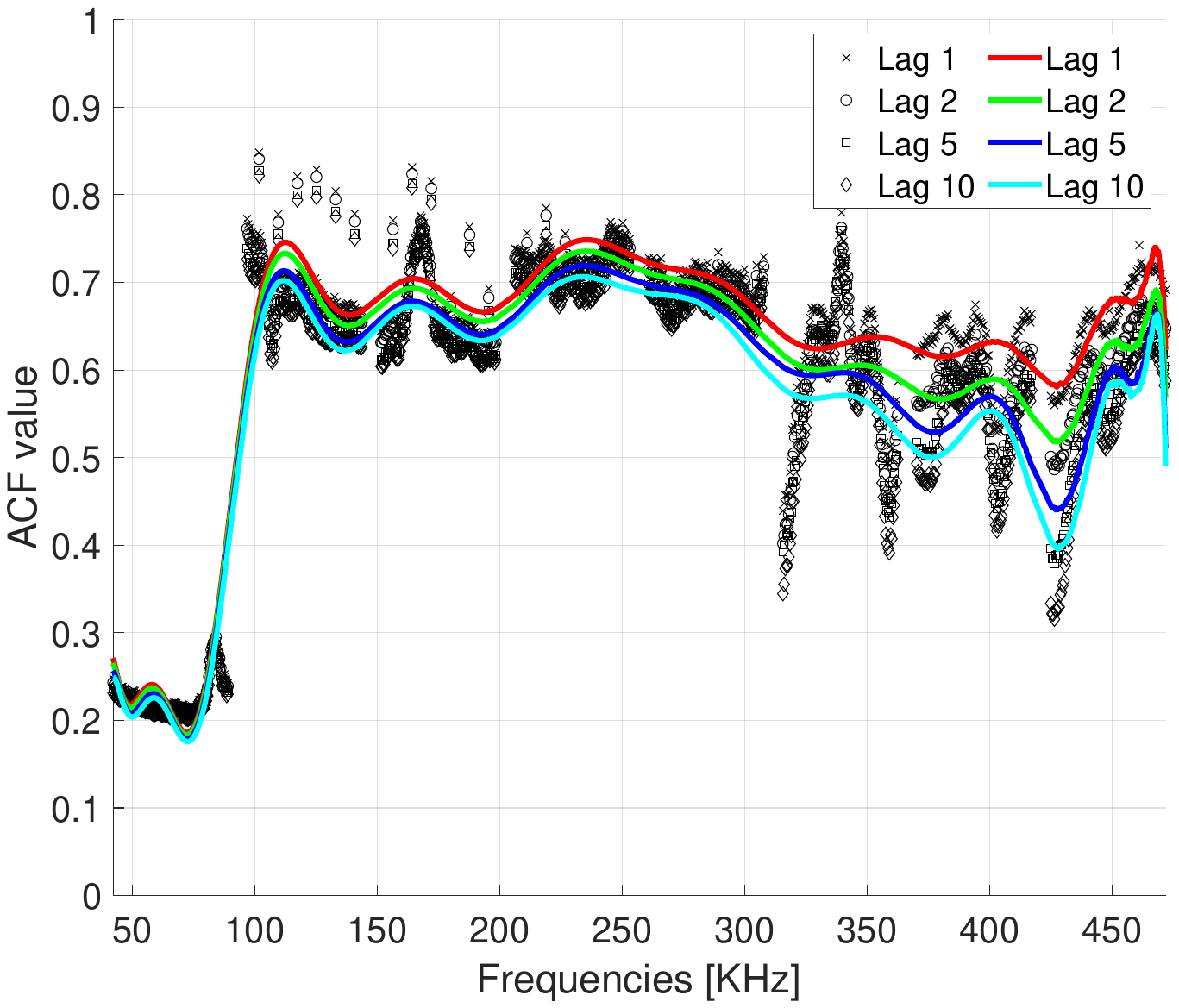}
        \subcaption{Location 2}\label{fig:lags_javier}
    \endminipage\hfill
    \minipage{0.33\textwidth}%
        \includegraphics[width=\linewidth]{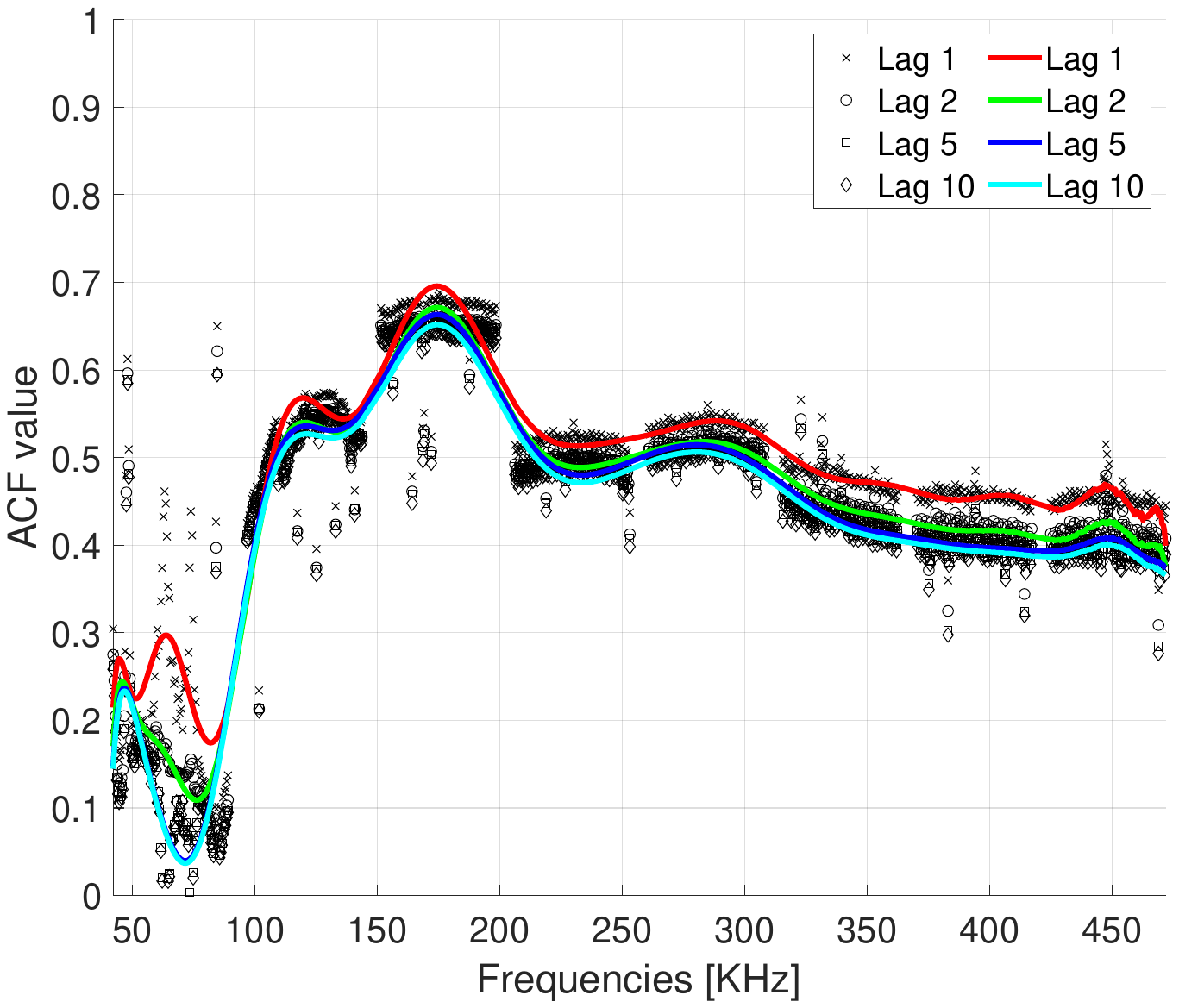}
        \subcaption{Location 3}\label{fig:lags_gabriele}
    \endminipage
    \centering
    \caption{Autocorrelation (\ac{ACF}) values at lags 1, 2, 5, and 10 as a function of the frequency for the three considered locations. We also consider the interpolations of the aforementioned values with the solid red, green, blue and cyan lines, respectively.}
    \label{fig:lags}
\end{figure*}

{\bf Independence test.} We began our analysis of the independence of the noise samples with the autocorrelation. We computed the \ac{ACF} up to lag 10 for all the frequencies and  considered locations. \ac{ACF} is the correlation of a signal with a delayed copy of itself as a function of the delay (i.e., lag). \ac{ACF} is an essential tool because it allows estimating whether the noise sample at time $k$ can be expressed (forecast) as a function of the values at previous lags. For each location, we computed the \ac{ACF} at lags 1, 2, 5, and 10, and we reported them as a function of the frequency in Fig.~\ref{fig:lags}. 

Figure~\ref{fig:lags} shows the autocorrelation function computed over a time span of 3600 lags (1 h). 

Firstly, we considered the significance bounds, computed from Bartlett’s approximation, which yielded:

$$ b = \pm \frac{z_{(1-\alpha)/2}}{\sqrt{N}}$$

where $z$ is the quantile of a normal distribution, and $\alpha=$0.95 and $N$ is the length of the trace, yielding $b = \pm$ 0.0022. Suppose the autocorrelation value is between the lower and the upper bounds. In that case, we can infer that there is no correlation at the considered lag and the null hypothesis of independence cannot be rejected at a significance level of $\alpha$.

In order to better highlight the behavior of adjacent frequencies, we interpolated the \ac{ACF} values at lags 1, 2, 5, and 10 with a polynomial fitting curve of high order (20), and we depicted the result of the interpolation using the solid red, green, blue, and cyan colors for lags 1, 2, 5 and 10, respectively. 

All the locations exhibited different levels of positive correlation for all the considered lags:

\begin{itemize}
    \item We observed that for all locations, the frequencies belonging to channel 1 showed the smallest correlation values, i.e., 0.1, 0.2, and 0.1 for locations 1, 2, and 3, respectively.
    \item The correlation was higher at lag 1, while it becomes smaller when the lags increased to a value 10.
    \item We observed that frequencies belonging to channels 2, 3, 4, and 5 were characterized by higher correlation values, i.e., approximately 0.5, 0.7, and 0.5, for location 1, 2, and 3, respectively.
\end{itemize}


Moreover, we checked the independence of the noise samples by resorting to the Ljung-Box Q-test, being defined as:

\begin{itemize}
    \item $H_0$: samples are independently distributed, 
    \item $H_1$: samples are not independently distributed.
\end{itemize}

while the test statistic is defined according to Eq.~\ref{eq:ljung-box_test}.

\begin{equation}
    Q(T, L, k) = T (T+2) \sum_{k=1}^L \frac{\rho(k)^2}{T-k}
    \label{eq:ljung-box_test}
\end{equation}

where $T$ is the sample size, $L$ is the number of autocorrelation lags, and finally, $\rho(k)$ is the autocorrelation value at lag $k$.
Under $H_0$, the statistic $Q(T, L, k)$ asymptotically follows a $\chi^2_L$ distribution with $L$ degrees of freedom, and therefore the null hypothesis can be rejected if $Q(T, L, k) > \chi^2_{1-\alpha, L}$, i.e., the statistic is strictly greater than the $1-\alpha$ quantile of the $\chi^2_L$ distribution, where $\alpha$ the level of significance. 

We evaluated $Q(T, L, k) > \chi^2_{1-\alpha, L}$, with $L=$ 10 s and $\alpha=$ 0.05, and the test confirmed that the null hypothesis ($H_0$) could be rejected for all the lags at all the frequencies. Therefore, we concluded that consecutive noise values (up to 10) were not independent while being correlated, as was also confirmed by the previous analysis in Fig.~\ref{fig:lags}.

\section{Modeling and Forecasting}
\label{sec:modeling}

\begin{figure}[h]
    \includegraphics[width=\columnwidth]{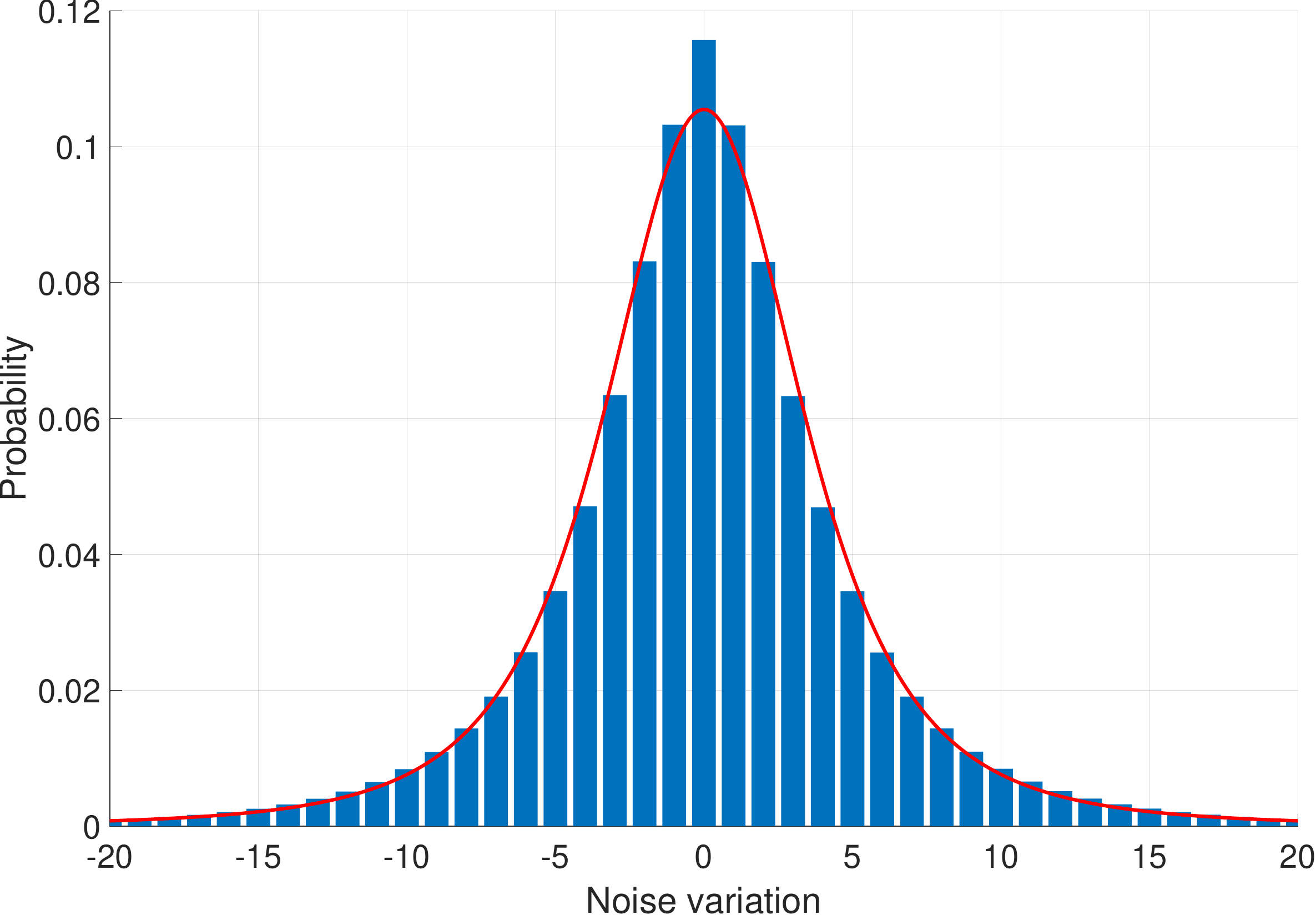}
    \centering
    \caption{Empirical distribution (blue bars) associated with the differentiate of two consecutive noise values $d(t) = n(t) - n(t-1)$, and the $t$ scale-distribution being the best-fit according to the maximum likelihood criteria.}
    \label{fig:dnoise}
\end{figure}

We considered the sequence of noise samples $n(t)$, and we computed the difference between consecutive values, i.e., $d(t) = n(t) - n(t-1)$, with $t > 1$. The blue bars in Fig.~\ref{fig:dnoise} show the probability distribution function associated with the noise variation $d(t)$ computed over consecutive noise samples considering all the frequencies and all the locations.

We performed a best-fit (maximum-likelihood) over the empirical distribution by considering several well-known distributions. We found that the best-fit was represented by the $t$ location-scale with a location parameter $\mu=$1.8$\cdot 10^{-3}$, scale parameter $\sigma=$3.47, and shape parameter $\nu=$ 2.87. The solid red line in Fig.~\ref{fig:dnoise} shows the $t$ location-scale function with the aforementioned parameters. 

Figure~\ref{fig:dnoise} proves that the vast majority of the noise samples kept the same value as the previous one (approximately 12\% of the cases), while more than 30\% of the samples is affected by minor variations, i.e., $\pm 1$. These findings confirm the results obtained by the autocorrelation (recall Fig.~\ref{fig:lags}), where we observed a significant correlation at lag 1 and 2. 

Our measurements, and the related analysis, prove that noise samples can be considered a stationary process, highly correlated, and very ``slow'', i.e., the noise process is characterized by a persistent mean value that is affected by small fluctuations only, i.e., less than $\pm 2 dB\mu V$ in 50\% of the cases.

\begin{figure*}[htbp]
    \minipage{0.33\textwidth}
        \includegraphics[width=\linewidth]{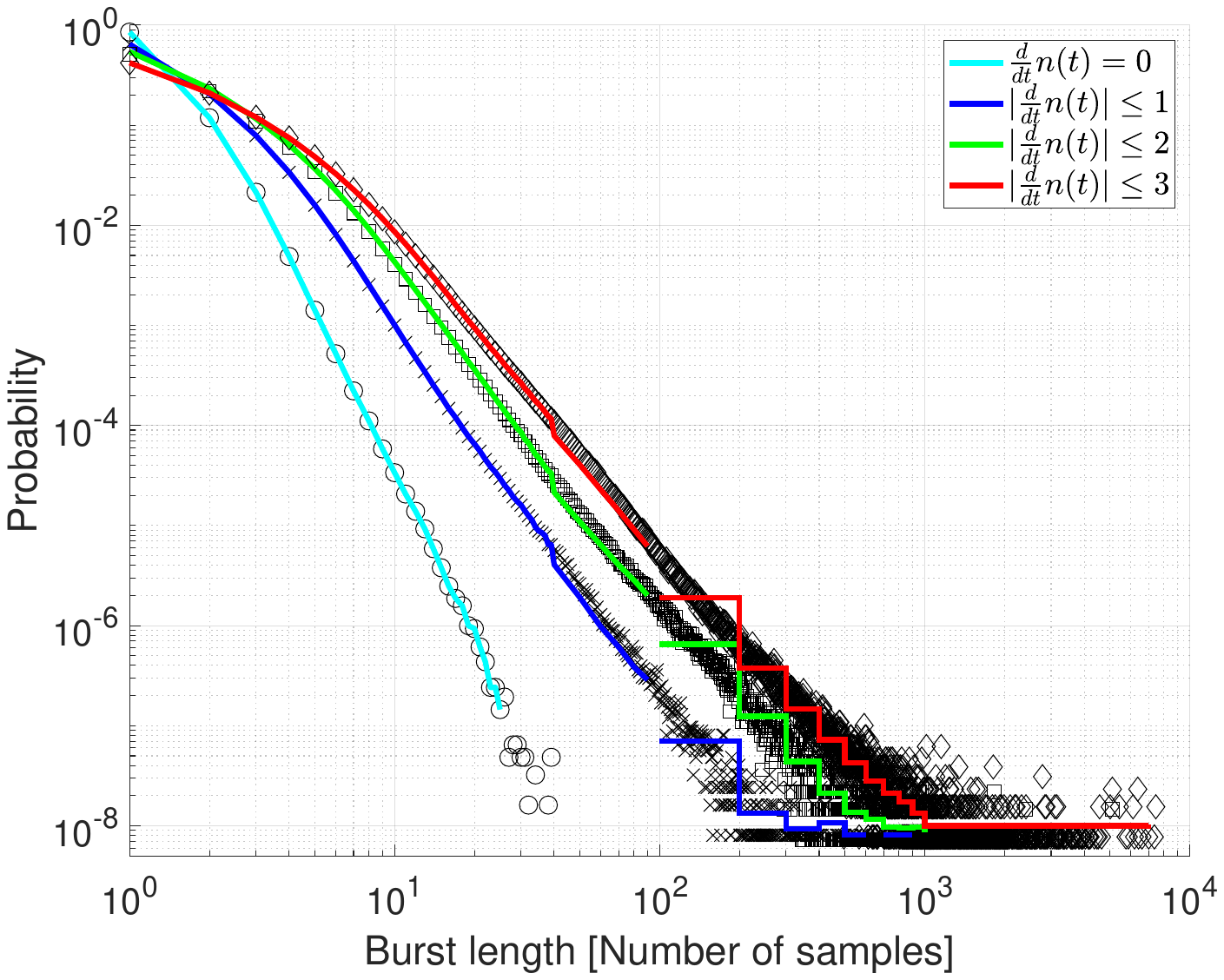}
        \subcaption{Location 1}\label{fig:burst_aymen}
    \endminipage\hfill
    \minipage{0.33\textwidth}
        \includegraphics[width=\linewidth]{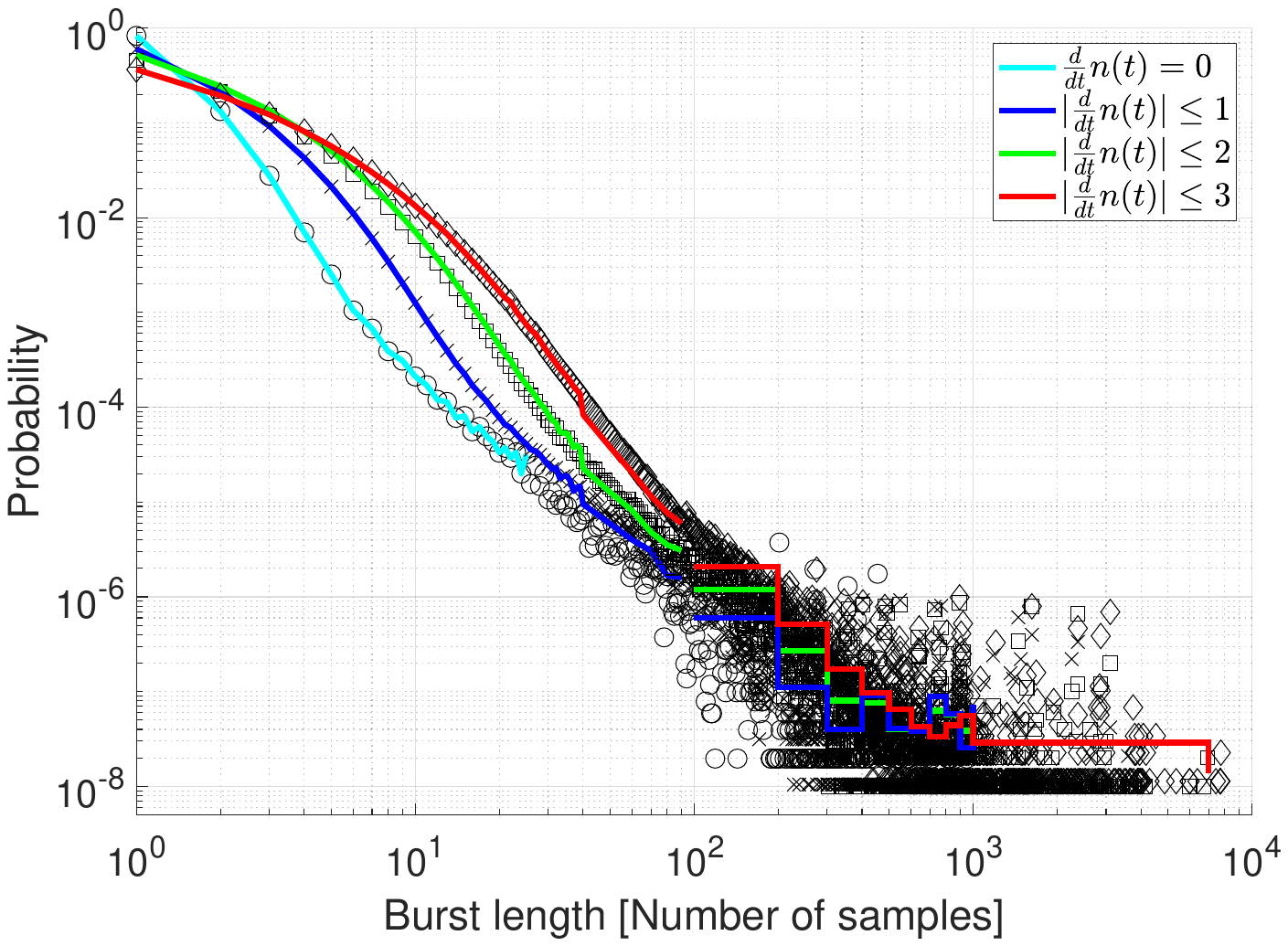}
        \subcaption{Location 2}\label{fig:burst_javier}
    \endminipage\hfill
    \minipage{0.33\textwidth}
        \includegraphics[width=\linewidth]{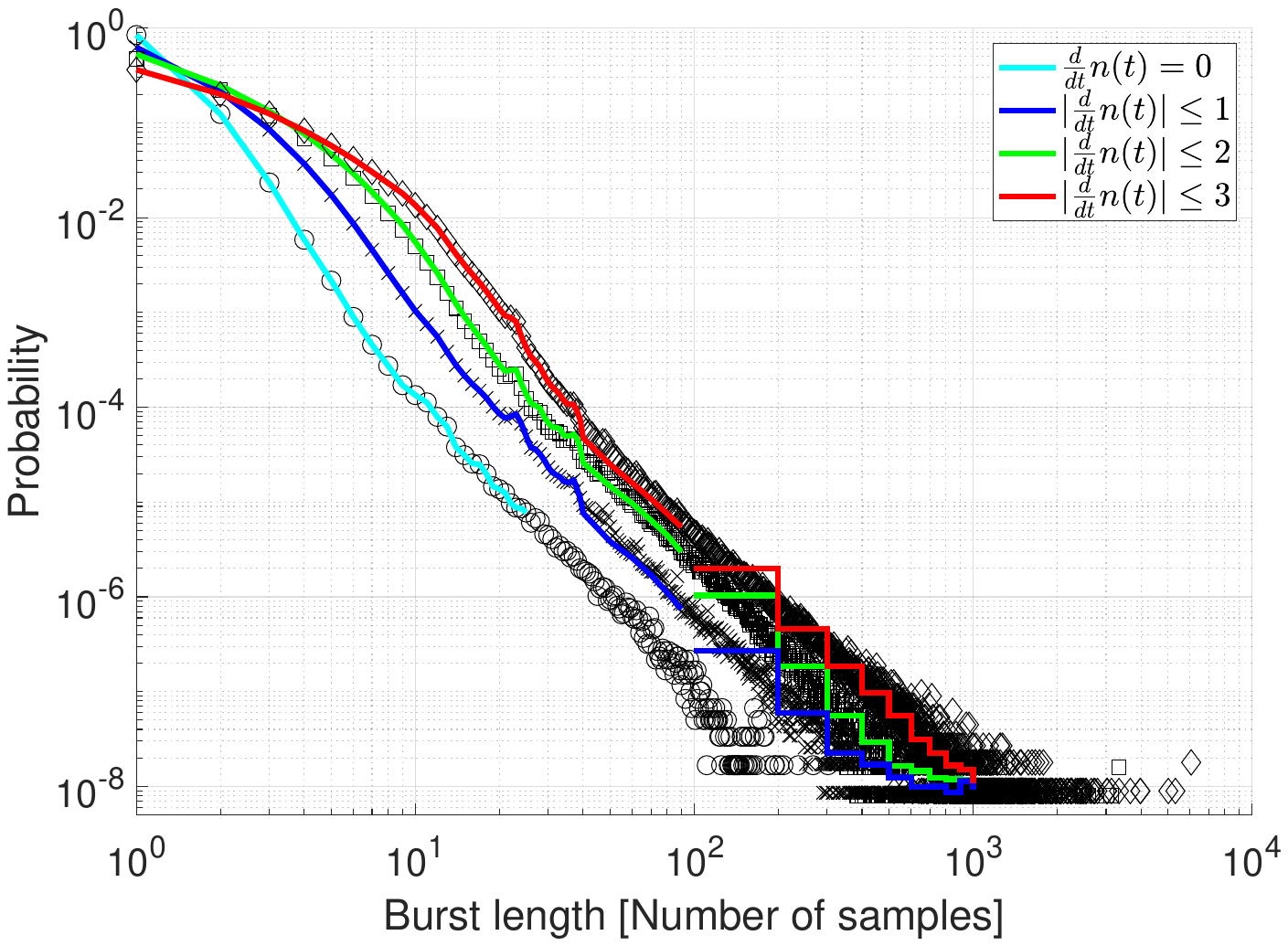}
        \subcaption{Location 3}\label{fig:burst_gabriele}
    \endminipage
    \centering
    \caption{Steady-state analysis of the noise process: we consider different derivative values $D \in \{0, 1, 2, 3\}$ and count for how many subsequent noise samples (bursts) feature $\frac{d}{dt}n(t) \le D$.}
    \label{fig:burst}
\end{figure*}

{\bf Steady-state analysis.} In the following, we consider the duration of steady-state (burst) periods in the sequences of the collected noise values. As for the previous analysis, for each location, we computed the derivative associated with all the noise samples (by considering the frequencies all together), and we investigated different levels of variations by considering the following cases:

\begin{itemize}
    \item $\frac{d}{dt}n(t) = 0$: Adjacent noise samples with the same value.
    \item $\frac{d}{dt}n(t) \le 1$: Adjacent noise samples with derivative values not exceeding 1.
    \item $\frac{d}{dt}n(t) \le 2$: Adjacent noise samples with derivative values not exceeding 2.
    \item $\frac{d}{dt}n(t) \le 3$: Adjacent noise samples with derivative values not exceeding 3.
\end{itemize}

The objective of this analysis is to investigate the frequencies of abrupt changes in the noise process and to assess the length of (almost) steady periods in the sequence of the noise values. Figure~\ref{fig:burst} shows the four cases mentioned above in the three locations. First, we observe that similar trends characterize the three locations. However, there are peculiarities due to the different number of occurrences, which might have been due to either the location or the superposition of all the frequencies. Moreover, we observe a strong linear relation starting at a burst length equal to 2 s up to a burst length equal to 100 s. Given the plot's log-log scale nature, we highlighted the geometric distribution nature of the process where bursts of length $k$ were constituted by independent $k$ events with derivative less or equal than $D$, with $D$ being 0, 1, 2, and 3. We note that the independence is related to the values of the derivative and not (as previously discussed) to the absolute values of the noise process, which are characterized by a strong correlation.

\section{Wrap-up, Comparison, and Discussion}
\label{sec:discussion}

In this section, we briefly recall the most significant findings of our contribution, we compare our statistics and models with the literature, and finally, we discuss their importance and impact.

\subsection{Wrap-up}

We considered several aspects associated with noise samples collected from the \ac{NB-PLC} frequency spectrum. The overall number of collected samples sums to 1,826,665,200, with 690,019,200, 516,466,800, and 620,179,200, collected at locations 1, 2, and 3, respectively. We firstly verified that the sampling period was consistent during the long measurement period (approximately 10 days) by considering the probability mass function (and the related cumulative probability) associated with the derivative of the sampling period (recall Fig.~\ref{fig:sampling_analysis}). 

We started our analysis by looking at the overall \ac{NB-PLC} spectrum, i.e., eight channels constituted of 776 frequencies in the range between 41,992 kHz and 471,679 kHz  (recall Fig.~\ref{fig:spectrum}). The baseline statistics (minimum, maximum, quantiles 10, 50, and 90) highlight the common patterns to the three locations and some significant differences. Indeed, the average noise level is higher in channel 1 independently of the location, and after an abrupt attenuation between channel 1 and 2, the noise level decreases toward the higher frequencies. It is worth noting that this phenomenon affects all the statistics mentioned above independently of the location. Conversely, each location is affected by peculiar noise patterns such as spikes, or higher/lower noise values at specific frequency values. We stress that our analysis does not consider other important parameters such as attenuation and grid impedance that might become significant at higher frequencies.

We then moved to evaluate the stationarity of the noise process (Fig.~\ref{fig:stationarity}). We searched the traces for chunks (of different sizes) having stationarity properties (testing negative to the unit root test). We verified that when the chunk length increased, the chunks' noise samples tended to be less stationary. Channels behave in the same way independently of the locations, apart from the anomaly of channel 1 in locations 2 and 3 (recall Fig.~\ref{fig:stationarity_javier} and Fig.~\ref{fig:stationarity_gabriele}). We noted that  a chunk length of 30 samples (lasting 30 s) has a probability of greater than 90\% of being stationary.

Moreover, we considered the ACF as a function of frequency and location. We evaluated the ACF at lags 1, 2, 5, and 10, and as for the previous analysis, we found common patterns at all the locations. While we observed that some channels could experience lower ACF values, i.e., channel 1, all the considered lags exhibit a high correlation at all the frequencies and independently of location. 

Finally, we considered a model for the noise process (Fig.~\ref{fig:dnoise}). We computed the derivative for all the noise samples and we searched for a best-fit according to several distributions: we evaluated the $t$-location scale as the  distribution having the best fit to our data. We observed that approximately 50\% of adjacent values exhibited a derivative value between -3 and 3, therefore, we analyzed further the distribution of consecutive samples, considering the samples that featured the same derivative values (Fig.~\ref{fig:burst}). We showed that consecutive noise samples were not affected by large fluctuations; indeed, approximately 50\% of our data $n(t)$ was characterized by long bursts with $\frac{d}{dt}n(t) \le 3$.

\subsection{Discussion}
In this section, we discuss the main results of our analysis. We consider two orthogonal directions: frequency and time.

{\bf Frequency.} The analysis of the noise in the NB-PLC spectrum shows a close relation independently of the considered locations. The noise levels become smaller when increasing the frequency. Moreover, we measured significantly higher noise levels for all the locations for frequencies lower than 100kHz, which becomes smaller (with different trends) depending on the location up to about 40 $dB\mu V$ at 400 kHz. Our measurements and the related analysis suggest leveraging the highest possible frequencies for the PLC protocols.

{\bf Time.} As for the time dimension, we focused on several aspects such as stationarity, correlation, independence, and forecasting. The noise process turns out to be stationary (up to 2 minutes of subsequent samples in more than 80\% of considered cases), i.e., the noise sequence keeps the same first-order statistics for a significant amount of time (up to 2 minutes). By combining the later observation with our remarks on the frequency domain, we suggest that it is better to move to higher frequencies despite waiting or implementing more robust error correction codes in the presence of a high level of noise. The correlation analysis confirms our previous claim: ACF experiences moderate to high correlation values up to lag 10. Nevertheless, we highlight the anomaly of a very low correlation (less than 0.1) for frequencies lower than 100 kHz. By combining these results with our observations from the frequency domain, our intuition is that low frequencies are affected by random (less correlated) noise. 
Finally, we focused on the possibility of forecasting the noise levels: we showed that subsequent noise values are in the range of $\pm 1$ in the 30\% of cases and they are not likely to be affected by abrupt changes. All our findings confirm a stationary noise process that tends to keep the same range of values for a long time.

{\bf Remarks and way ahead.} This paper sheds light on several aspects affecting the noise levels in residential houses that should be considered when designing a PLC protocol working at the NB-PLC frequencies. We showed that the noise affects the NB-PLC spectrum in different ways (as a function of the frequency) and it is characterized by strong stationarity and correlation patterns. We are confident that our findings, including the statistics and forecasting models, will help to design better PLC protocols with improved performance. Although being as general as possible, our analysis focuses only on residential housing, opening the way for similar analysis in industrial and commercial scenarios.

\section{Conclusion}
\label{sec:conclusion}

We introduced a statistical description of the noise process in the \ac{NB-PLC} frequency band. Our data collection involved more than 1.8 billion samples taken from three different locations over a time span of approximately 10 days. Our analysis focused on the noise time series from the NB-PLC bandwidth. First, we found a correlation (up to 30 s) at the intra-frequency level. Moreover, we provided a thorough analysis involving stationarity, autocorrelation, and independence. Finally, we proved that the derivative of the noise process could be modeled according to a $t$-location scale distribution, while being characterized by a sequence of values affected by small fluctuations, in the range of $3 dB \mu V$.

\section*{Acknowledgment}
This publication was made possible by NPRP grant NPRP12S-0125-190013 and NPRP12C-0814-190012 from the Qatar National Research Fund (a member of Qatar Foundation). The findings achieved herein are solely the responsibility of the authors.
\bibliographystyle{IEEEtran}
\balance
\bibliography{PLCNoiseModeling_R2}

\end{document}